\documentclass[10pt,letterpaper,man,floatsintext]{apa7}
\usepackage{epsfig}
\usepackage{lineno}
\usepackage{graphics}
\usepackage{times}
\usepackage{epsfig}
\usepackage{graphicx}
\usepackage{amsmath}
\usepackage{array}
\usepackage{textcomp}
\usepackage{setspace}
\usepackage{afterpage}
\usepackage{hyperref}
\usepackage{xurl}
\usepackage[natbibapa]{apacite}

\newcommand{\captionletter}[1]{{#1}}

\title{Comparing Constellations across Cultures} 
\shorttitle{Comparing Constellations across Cultures} 

\authorsnames[1,2,1,1]{Charles Kemp, Duane W.\ Hamacher, Daniel R.\ Little, Simon J.\ Cropper} 
\authorsaffiliations{{Melbourne School of Psychological Sciences, University of Melbourne},{School of Physics, University of Melbourne}}

\abstract{
Cultural astronomy reveals ways in which perception and culture have shaped the interpretation of the night sky.
}

\authornote{\addORCIDlink{Charles Kemp}{0000-0001-9683-8737}\\ 
\addORCIDlink{Duane W.\ Hamacher}{0000-0003-3072-8468}\\ 
\addORCIDlink{Daniel R.\ Little}{0000-0003-3607-5525}\\
\addORCIDlink{Simon J.\ Cropper}{0000-0002-3574-6414}

We thank Doina Bucur for comments on the manuscript.
Correspondence and request for materials should be addressed to Charles Kemp, 
Melbourne School of Psychological Sciences, University of Melbourne, Victoria 3010, Australia.  E-mail: \texttt{c.kemp@unimelb.edu.au}

}

\begin{document}
\maketitle

John Herschel referred to the Western constellations as ``uncouth figures and outlines of men and monsters,'' and claimed that astronomers ``treat them lightly, or altogether disregard them, except for briefly naming remarkable stars, as $\alpha$ Leonis, $\beta$ Scorpii, \&c.\ \&c., by letters of the Greek alphabet attached to them'' \citep[p 156]{herschel1836}.
His contemporary Thomas Dick was even more scathing, and wrote that these ``fantastical groups ... may comport with the degrading arts of the astrologer,'' but are ``not only incompetent to the purposes, but completely repugnant to the noble elevation of modern astronomical science''\citep[p 46]{dick1844}. 

\begin{figure}[t]
\begin{center}
% had to crop manually
\includegraphics[width=\textwidth]{{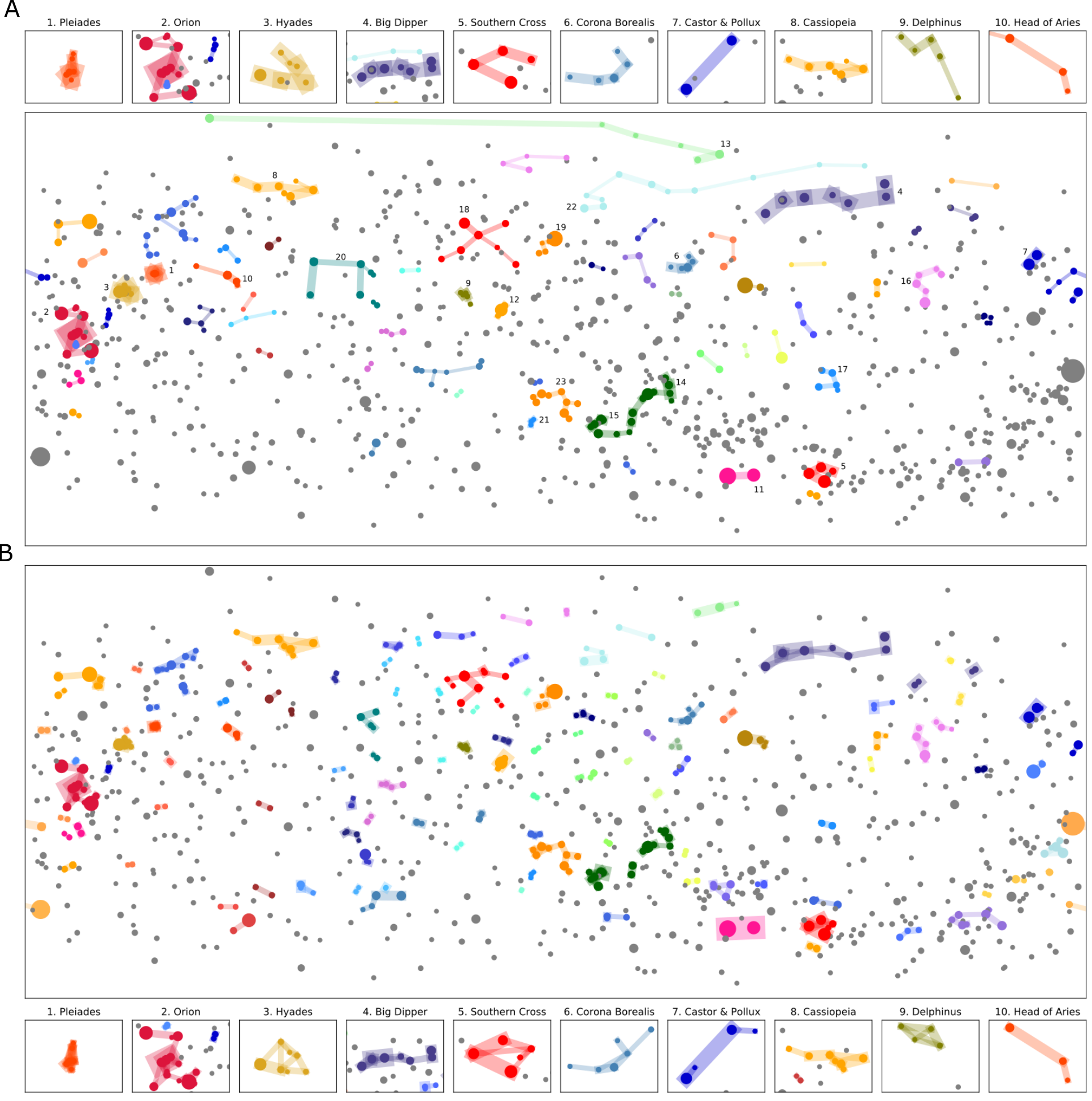}}
\caption{Common star groups across cultures compared with model predictions. (\captionletter{A.}) Consensus system created by overlaying minimum spanning trees for all star groups in a data set of 27 sky cultures. Edge widths indicate the number of times an edge appears across the entire dataset, and edges that appear three or fewer times are not shown. Node sizes indicate apparent star magnitudes, and only stars with magnitudes brighter than 4.5 have been included.  Insets show 10 of the most common asterisms across cultures, and numbers greater than 10 identify the Southern Pointers (11), shaft of Aquila (12), Little Dipper (13), head of Scorpius (14), stinger of Scorpius (15), sickle in Leo (16), Corvus (17), Northern Cross (18), Lyra (19), Square of Pegasus (20), Corona Australis (21), head of Draco (22) and the teapot in Sagittarius (23). 
 (\captionletter{B.}) Star groups according to a model of perceptual grouping. The edges shown belong to minimum spanning trees of groups identified by the model, and edge widths are proportional to strengths assigned by the model. Reprinted from \citet{kemphlc20}.}
\label{fig:fullsky}
\end{center}
\end{figure}
\afterpage{\clearpage}

Other scholars, however, treasure constellations for the window that they provide onto ancient and modern cultures~\citep{krupp91,aveni19}. 
%cultures~\cite{krupp91}. 
Over thousands of years, researchers including anthropologists and astronomers have documented the many different ways in which cultures around the world organize the night sky into systems of stories.  In recent years, the planetarium software Stellarium~\citep{zottihwcc} has provided a convenient platform for documenting and sharing these ``sky cultures'', and the resulting data
can be used to compare constellations across dozens of cultures.

Here we discuss the different ways in which stars have been organized into groups and in which these groups have been endowed with meaning.   Psychologists have studied how the human perceptual system organizes simple visual elements such as dots or contour fragments into groups~\citep{wagemans12I,wagemans12II,elder15},  and within this literature constellation formation is often invoked as an example of perceptual grouping~\citep{kohler29,metzger36}. 
To a good first approximation, the human visual system is invariant across cultures and therefore offers up similar candidate star groups to any two people observing the same region of the night sky. Given this foundation, culture then shapes which groups attract the shared attention of a community and the ways in which these groups are embedded in systems of stories. 

\section{Star groups}

If constellations are shaped by universal perceptual principles, then convergences in the star groups identified by different cultures should be expected. 
\citet[p 58]{krupp00} identifies a ``narrow company'' of constellations that are common across cultures, including the Pleiades, Orion's Belt, the Big Dipper, and the Southern Cross, but states that ``after that, the picture is very muddy''\citep{krupp_maat}. After the small set of near-universal constellations, there is presumably a long tail of constellations that recur across cultures to different extents, but characterizing this gradient is impossible without a relatively large database of groupings across cultures. 

We recently assembled such a database by drawing in part on the set of Stellarium sky cultures~\citep{kemphlc20}. Figure \ref{fig:fullsky}A shows the groupings that appear most frequently across the 27 sky cultures included in the database. As expected, Krupp's ``narrow company'' of near-universal constellations is strongly present in the data, but the plot also highlights groupings such as Corona Borealis, Delphinus, Corona Australis and Corvus that are discussed less often but nevertheless identified by multiple cultures.

To test the idea that universal perceptual principles account for many of the groups that recur across cultures, we used a simple model of perceptual grouping to organize stars into groups. The model considers only brightness and proximity, and preferentially groups nearby pairs of stars where the fainter of the pair is still relatively bright. Figure \ref{fig:fullsky}B shows that the model partially captures many of the groups that recur across cultures, including examples such as Corona Borealis that go beyond Krupp's ``narrow company'' of near-universal constellations. The results therefore provide some initial evidence that perceptual grouping is sufficient to explain a substantial proportion of the star groups found across cultures. 

\begin{figure}
\begin{center}
% had to crop manually
%\includegraphics[width=\textwidth]{{figures/fig2_crb.pdf}}
\includegraphics[width=0.75\textwidth]{{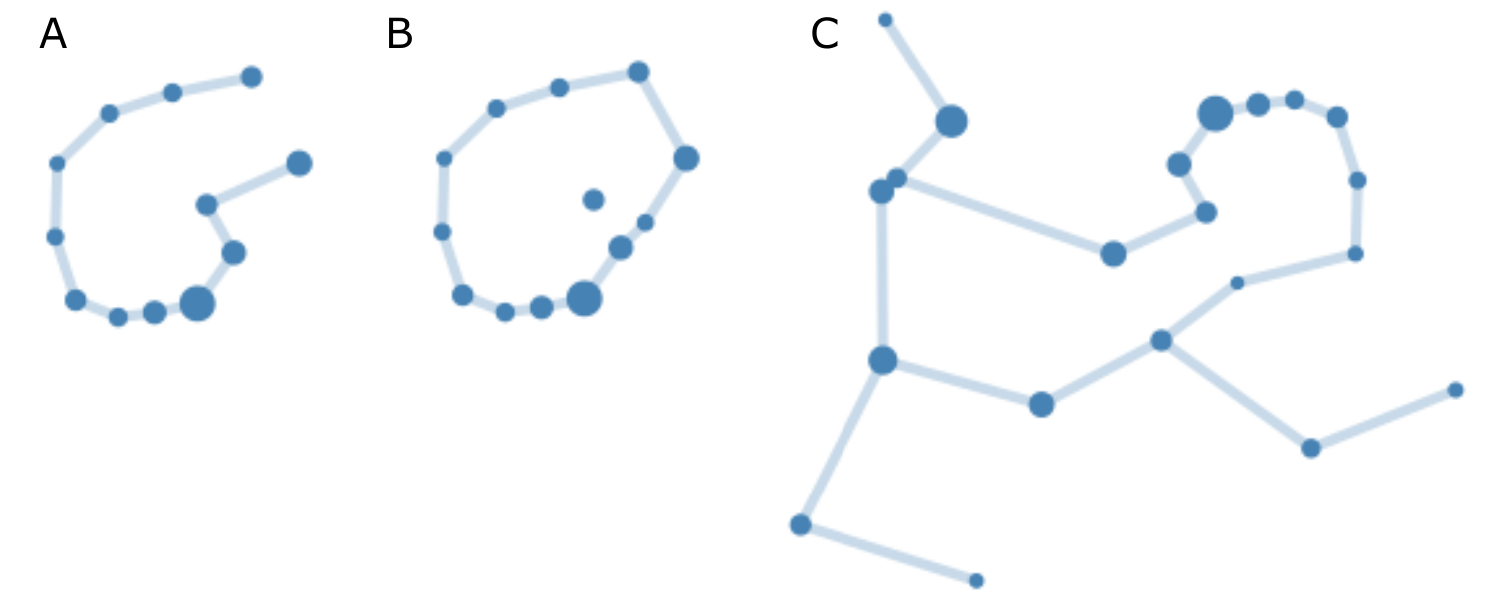}}
%\includesvg{figures/na_combinedplain.svg}
\caption{Three constellation figures that include the brightest stars of Corona Borealis (CrB). \captionletter{A.} The Micmac see CrB as the den of a bear, which is represented by the bowl of the Big Dipper. \captionletter{B.} The Pawnee see CrB as a Council of Chiefs, and the central star ($\theta$ CrB) has been described as an errand man or as a servant cooking over a fire.
\captionletter{C.} The Kalina see CrB as the head of Kaitusiyuman, the constellation of the jaguar. Panel \captionletter{C} is inverted relative to panels \captionletter{A} and \captionletter{B} because the Kalina view Kaitusiyuman from the southern hemisphere. As is typical, there is some uncertainty about the star groups and figures shown in all three panels, and these elements often vary within cultures. For example, a Pawnee sky map~\citep{buckstaff27} shows 11 stars in the Council of Chiefs rather than the 12 or 13 shown here, and some Kalina locate Kaitusiyuman near Andromeda.  Panels \captionletter{A} and \captionletter{B} are based on \protect\citet[p 37,223]{miller97}, and \captionletter{C} is based on \protect\citet[p 130]{maganaj82}.}
%Maga\~{n}a and Jara 
\label{fig:crb}
\end{center}
\end{figure}

In addition to picking out prominent groups of stars, some cultures ``connect the dots'' and organize these stars into constellation figures.  Three examples that include the brightest stars in Corona Borealis are shown in Figure~\ref{fig:crb}. The first two examples are similar, but the third is rather different and includes parts of the western constellation of Bo\"{o}tes. Like the star groups themselves, figures defined over these groups appear to be influenced by perceptual principles, including the principle that figures which minimize the sum of all edge lengths tend to be preferred~\citep{dry09}. A recent study compared constellation figures across a set of 50 sky cultures, and among other analyses ranked a set of bright stars based on the diversity of the figures to which they belong~\citep{bucur21}.  The results revealed that constellation figures including $\beta$ and $\delta$ Corona Borealis showed relatively low diversity, and the greatest diversity was found for figures including Betelgeuse ($\alpha$ Ori). 

%computed the variability across figures that include a given star~\cite{bucur21}. 

%The brightest stars in Corona Borealis tend to be organized into either a chain or a ring, which means that constellation figures for these stars are mostly consistent across cultures. In contrast, the analysis reveals that constellation figures including Betelgeuse  ($\alpha$ Ori) vary widely across cultures.

\section{Constellation stories}

After identifying prominent groups of stars, a community must decide how these groups are named and what meanings they may have.  \citet[p 4]{herschel1841} notes that ``few tolerable resemblances to objects of ordinary terrestrial occurrence can be made out among the stars'', and as a result, constellation names are influenced more by culture than by visual perception.  In many cases, constellations are embedded in stories that provide insight into central aspects of culture, including knowledge about plants, animals, and seasons in addition to sacred traditions and ceremonies~\citep{hamacher22}.  Comparing these stories across cultures therefore highlights the values, interests, and priorities of different cultures. 

Constellation stories for many cultures have been compiled~\citep{allen99,staal88,krupp91}, but comparisons of stories across cultures often focus on Krupp's ``narrow company'' of constellations. One notable line of work considers the ``cosmic hunt,'' a story found in various forms across Europe, Asia, and the 
Americas~\citep{gibbon64,berezkin05}.  In a common version of the story, three stars in the handle of the Big Dipper are hunters, and the bowl of the Big Dipper is a bear that the men are pursuing. The geographic distribution of this story provides evidence about cultural influence in the distant past and suggests that the first settlers of North America may have brought the bear story with them 14,000 years ago when they crossed the Bering Strait.

%\afterpage{\clearpage}

The shared groupings that emerge in Figure \ref{fig:fullsky}A suggest other possible targets for comparisons of constellation stories across cultures, including Corona Borealis (CrB) and Corona Australis (CrA). In a striking convergence, both groups
 were seen as wreaths by the Greeks and by the Micronesian people of the Marshall islands~\citep{erdland14}. 
Multiple cultures saw these groups as containers including a basket (CrB, Hawaiian~\citep{chauvin00}), bowl or platter (CrB and CrA, Arab~\citep{allen99}), coolamon (CrA, Western Arrernte~\citep{hamacherg13}), eagle's nest (CrB, Wiradjuri~\citep{leamanh19}) and ostrich's nest (CrA, Arab~\citep{allen99}). Other recorded interpretations include enclosures such as a prison (CrB, Chinese~\citep{sun00}), sweat lodge (CrB, Ojibwe~\citep{lee14}), llama corral (CrA, Pacariqtambo~\citep{urton05}) and bear's den (CrB, Micmac~\citep{miller97});  creatures such as a a tortoise (CrB, Shipibo~\citep{roe05} and CrA, Chinese~\citep{sun97}), armadillo (CrB, Barasana~\citep{hughjones82}), crab (CrB, Arawak~\citep{jara05}), kookaburra (CrA, Wiradjuri~\citep{leamanh19}), and poisonous snake (CrA, Barasano~\citep{hughjones82}); and groups of people such as a council of chiefs (CrB, Pawnee~\citep{chamberlain92}) or a group of people sitting around a fire (CrA, Khoisan~\citep{lloyd}).
% Multiple cultures saw these groups as containers including a basket (CrB, Hawaiian), bowl or platter (CrB and CrA, Arab), coolamon (CrA, Western Arrernte), eagle's nest (CrB, Wiradjuri) and ostrich's nest (CrA, Arab). Other recorded interpretations include enclosures such as a prison (CrB, Chinese), sweat lodge (CrB, Ojibwe), llama corral (CrA, Pacariqtambo) and bear's den (CrB, Micmac);  creatures such as a a tortoise (CrB, Shipibo and CrA, Chinese), armadillo (CrB, Barasana), crab (CrB, Arawak), kookaburra (CrA, Wiradjuri), and poisonous snake (CrA, Barasano); and groups of people such as a council of chiefs (CrB, Pawnee) or a group of people sitting around a fire (CrA, Khoisan).

Understanding any one of these interpretations typically requires learning how it is embedded in an entire system of knowledge and beliefs.  In the Marshallese sky culture, CrB (``Wreath of the Ijjirik") and CrA (``Wreath of the Erribra") are set in opposition to each other. The Ijjirik and the Erribra are clans that live in different parts of the Marshall islands, and the two celestial wreaths are linked with perceived characteristics of these clans. According to \citet{erdland14}, the Ijjirik are fickle and never agree with each other, which is why the stars of CrB are unevenly spaced and vary in brightness. In contrast, the Erribra are united and do not change their minds, which is why the stars of CrA form a regular semicircle and are near constant in brightness.

For the Pawnee of North America, CrB is known as the ``Council of Chiefs'' and portrays ``great mythical councils where creation was planned, then carried out, as well as councils to be held on Earth"\citep{chamberlain96}.  The Pawnee ``trace their origin and organization to the stars''~\citep[p 10]{fletcher03}, and associated stars with their yearly cycle of religious ceremonies and with activities such as horticulture, buffalo hunting, and war~\citep{murie81}. 
The Council of Chiefs derives its meaning from its place within this intricate system of  knowledge~\citep{chamberlain82,chamberlain92}, 
but here a single story will have to suffice. In this story,  a ``poor, lame, one-eyed boy'' is magically brought before the Council of Chiefs, who heal his ailments, instruct him to go on the warpath, and promise that he will win himself a wife if he follows their instructions~\citep{dorsey04}. The boy, however, disregards their commands and is therefore transformed back to his former lame self. This story portrays the Council of Chiefs as a source of wisdom and authority and warns of the consequences of disobeying their commands. 

Just as computational methods have been used to analyze star groups~\citep{kemphlc20} and constellation figures~\citep{bucur21} across cultures, computational analyses of star stories have also been developed.  Berezkin pioneered this approach by compiling a catalogue of world mythology and folklore~\citep{berezkin15} that draws on data from around a thousand cultures and that includes stories about stars and other celestial objects in addition to stories about many other aspects of culture. Among other applications, these data have been used in a study that applied phylogenetic methods to analyze the geographic distribution of stories about Orion and the Pleiades~\citep{dhuy17}. Future work can aim to build on this foundation by supplementing Berezkin's data with stories about a broader set of constellations and by drawing on algorithms for natural language processing in order to compare star stories across cultures. 

\section{Conclusion}

We have suggested that comparative analysis of constellations and constellation stories can reveal general principles that shape the way in which people organize and interpret the night sky.  
%We focused here on star groups and their meanings, but names and stories associated with single stars~\cite{allen99},  dark sky constellations~\cite{gullberg20}, the sun, moon and planets,  and transient phenomena such as comets, novae and eclipses~\cite{hamachern11} can also be compared across cultures. 
We focused here on star groups and their meanings, but names and stories associated with single stars,  dark sky constellations, the sun, moon and planets,  and transient phenomena such as comets, novae and eclipses can also be compared across cultures. 
This work is intrinsically interdisciplinary and requires contributions from astronomers, anthropologists, ethnologists, historians, and custodians of traditional knowledge. Had he been so inclined, Herschel could have contributed to this effort by studying Indigenous sky cultures during the four years that he spent in South Africa. During this period, Herschel began to develop a proposal for reforming the Western constellations~\citep{herschel1841}, but helping to chart the diversity of constellations across cultures may ultimately have been a more valuable contribution. 

\bibliographystyle{apacite}
%\bibliography{../stars}

\end{document}